\documentclass[aps, llncs, preprint, nofootinbib]{revtex4-1}

\usepackage{amsfonts}
\usepackage{amssymb}
\usepackage{graphicx}
\usepackage{amsmath}
\usepackage{float}
\usepackage{color}
\usepackage{soul}
\usepackage{lineno}
\usepackage{siunitx}
\usepackage{physics}
\usepackage[utf8]{inputenc}
\usepackage[T1]{fontenc}
\usepackage{gensymb}
\DeclareUnicodeCharacter{0096}{_}
\usepackage{multirow}
\makeatletter
\newcommand{\printfnsymbol}[1]{%
  \textsuperscript{\@fnsymbol{#1}}}
\makeatother
\begin{document}

\title{Topological polar textures on CsPbBr$_{3}$ nanoplatelets}

\author{Monika Bhakar$^{1*}$, Pooja Bhardwaj$^3$, Gokul M.Anilkumar$^2$, Atikur Rahman$^2$, Goutam Sheet$^1$}
\email{monikabhakar37@gmail.com}
\email{goutam@iisermohali.ac.in}

\affiliation{$^1$Department of Physical Sciences, Indian Institute of Science Education and Research (IISER) Mohali, Sector 81, S. A. S. Nagar, Manauli, PO 140306, India}

\affiliation{$^2$Department of Physics, Indian Institute of Science Education and Research, Dr. Homi Bhaba Road, Pune, 410008, India}

\affiliation{$^3$University Center for Research and Development (UCRD), Chandigarh University, Punjab- 140413, India}

\begin{abstract}
Polar topological textures like the bubble domains, flux-closures, and labyrinth etc. unlock functional responses in ferroic systems but are difficult to stabilize and control in chemically simple, solution-grown materials. Here we show that ultra-thin, large-area CsPbBr$_3$ nanoplatelets host room-temperature ferroelectric bubble domains whose characteristic size is tunable by thickness. Using contact-resonance piezoresponse force microscopy (PFM) across 125~nm–2~$\mu$m, we observe a systematic decrease in domain size with decreasing thickness, consistent with a depolarization-field-controlled stability window. Repeated scanning transforms bubbles into labyrinthine patterns, indicating metastability under weak mechanical/electrical perturbations. Upon heating, bubbles evolve into labyrinths and vanish at $T_C\!\approx\!90^\circ$C, with domain nucleation recovered on cooling. These results establish a controllable platform for polar topology in a stable, stochiometric perovskite grown via a solvothermal route, and clarify how electrical boundary conditions (set by thickness and temperature) govern texture selection. The thickness-tunable polar textures identified here offer a route to engineer domain-wall–mediated functionalities in halide perovskites.
\\\\
\end{abstract}

\maketitle

\section*{Introduction}
In ferroic materials, the domain walls separating regions with differently oriented polarization may display novel topological structures like flux closures, vortices, bubbles, and Skyrmions \cite{Michael, Chun, A. K., Zhang1, Yu}. The potentially fascinating relationship and distinction between Skyrmions and larger size bubble domains motivated search for bubble domains in both ferromagnets and ferroelectrics. While such bubbles are relatively easily found in magnetic structures, their realization in ferroelectrics remains to be a challenging task. A ferroelectric bubble is a localized region of electric polarization in a ferroelectric material, where the polarization direction is reversed relative to its surrounding matrix.

Electric polar textures, analogous to magnetic skyrmions and bubbles, are stabilized through distinct mechanisms. In magnetic systems, these topological defects are stabilized primarily by long-range dipolar interactions. In contrast, electric polar textures are predominantly stabilized by the complex interplay of mechanical and electrical boundary conditions\cite{Lai, Kornev, Le, Nina, A., Vivasha}.
These boundary conditions are modulated by factors such as the depolarization field (E$_d$), externally applied electric fields, and mechanical stress or strain \cite{Lai, Xingchen}. Thin films provide enhanced control over ferroelectric properties and the depolarization field, enabling the tuning of these domain structures \cite{Zhang1}. In ultrathin films, the formation and stability of such polar textures are influenced by various parameters including film thickness, external electric field, and temperature\cite{Lai, Kornev}. Such polar structures have previously been stabilized by integrating perovskite thin films with other materials through the creation of heterostructures and superlattices\cite{Zhang1, Y., Tan, Jie, S}. Here, we report the stabilization of such rare topological structures in thin nanoplatelets of the perovskite CsPbBr$_{3}$ where enhanced photovoltaic properties were recently reported. 

CsPbBr$_3$ is a semiconductor with a wide bandgap $\sim$ 1.44 to 1.87 eV \cite{Raouia Ben Sadok}. Nanocrystals and nanodots of CsPbBr$_3$ have been shown to undergo a structural transition, exhibiting ferroelectric-like behavior\cite{S. Hirotsu, Ghosh, Xia }. In a recent study, we showed that thin nanoplatelets of CsPbBr$_3$ also host a ferroelectric transition for a thickness of 310 nm, where labyrinth or stripy domains stabilized\cite{Gokul}. First-principles calculations suggested that the depolarization field plays a critical role in the formation of the stripy configuration. Such calculations also investigated that the emergence of such stripes from a single ferroelectric domain structure should happen through intermediate domain structures in the diverse topological textures including the bubble domains\cite{Kornev}. Notably, E$_d$, and consequently the domain size in such ultrathin systems, can be controlled by tuning the thickness of the material\cite{Weng, Celine}. This insight into the modulation of domain structures via control of the depolarization field provides a strong motivation for detailed investigations into the ferroelectric domain structures in thin nanoplatelets of CsPbBr$_3$.

In this manuscript, we report the emergence of polar topological textures in CsPbBr$_3$ ultrathin nanoplatelets. Our Piezoresponse Force Microscopy (PFM) experiments revealed the formation of bubble domains ranging in size from 200 nm to 2 µm and their evolution by changing the thickness of the nanoplatelets. Time-dependent PFM investigations show that the polar bubble domains are easily transformed into labyrinth structures under repeated scanning of the PFM tip. This suggests that these structures are metastable and they appear only within a very narrow range of boundary conditions. The interaction with the PFM cantilever acts like a perturbation triggering the change in the domain pattern. Additionally, temperature-dependent studies demonstrated a reversible transition between bubble domains and the labyrinthine domain structure. As the nanoplatelet thickness is increased, E$_{d}$ diminishes, leading to a corresponding increase in domain width. Our results collectively suggest that the electrical boundary conditions, which govern the formation and stability of bubble domains, are finely tuned by variations in both thickness and temperature in this 
quasi two dimensional (2D) ferroelectric system.
 
\section*{Experiments and Results}
CsPbBr$_{3}$ nanosheets were synthesized on ITO-coated glass substrates using a solvothermal method reported elsewhere \cite {Gokul}. Atomic force microscopy (AFM) experiments were performed to determine the thickness of the nanoplatelets. Following that, thickness-dependent PFM measurements were carried out to examine the ferroelectric domain structure of the system as a function of the thickness. The formation and evolution of ferroelectric domains were probed through temperature and time-dependent PFM studies. In our PFM setup, as illustrated in Figure 1(a), a conductive tip is brought in contact with the surface of the sample. A laser beam is allowed to fall on the end of the back side of the cantilever. A sinusoidal AC voltage ($V_{ac}$) is applied to a piezoelectric chip attached to the cantilever to induce oscillation.  The cantilever is then tuned to its contact-mode resonance frequency, and all measurements are performed at this resonance frequency to get maximum sensitivity\cite{Leena, Jagmeet, Shubhra}. In the contact mode, the deflection of the cantilever is recorded as it scans over the surface of the sample. PFM measures the mechanical response of a sample to an electric field applied between the tip and the sample. The applied voltage induces a local deformation in the sample, either expansion or contraction, which alters the cantilever's deflection. This deflection of the cantilever is interpreted as the piezoresponse of the sample. 

To investigate the ferroelectric nature of domains and the coercive voltage as a function of film thickness, thickness-dependent Switching Spectroscopy Piezoresponse Force Microscopy (SS-PFM) measurements were performed. In SS-PFM, a combination of AC and DC bias is applied to the probe tip, inducing piezoelectric expansion or contraction in the material, depending on the polarity of the applied tip voltage. As the DC voltage is swept, this results in a characteristic switching of polarization, observable as hysteresis in the phase versus DC voltage ($\phi$ vs $V_{dc}$) curve. The deformation amplitude is observed as a butterfly loop in the $A_{\omega}$ vs. $V_{dc}$ curve. It is important to note that hysteretic behavior can arise from factors other than piezoelectricity, such as electrostatic or electrochemical effects\cite{Jesse1, Jesse2}. To minimize electrostatic contributions, a sequence of DC voltages with AC modulation is applied in a triangular sawtooth pattern. This switching spectroscopy method allows the measurement of the response signal in the "off" state of the applied pulses. Follow-up topographic analysis ruled out any contribution from local electrochemical reaction.
 \begin {figure}
 \includegraphics[height=5.5in, width=6.1in]{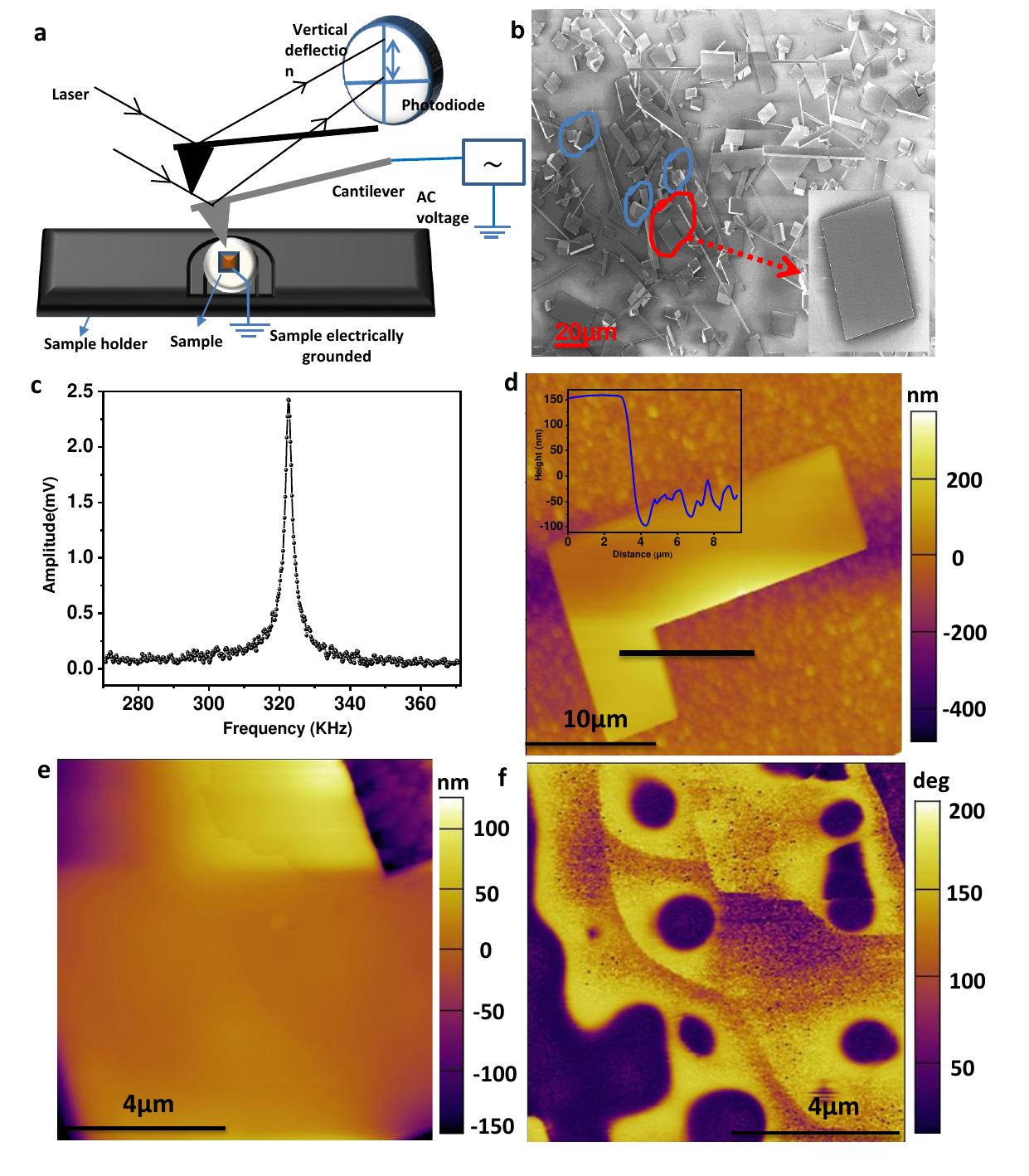}
 \caption{(a) Schematic representation of the Piezoresponse Force Microscopy (PFM) setup used for the measurements. (b) The field emission scanning electron microscopy (SEM) image shows CsPbBr$_{3}$ sheets of varying thickness. Sheets delineated by red regions are identified as thin, while those highlighted in blue represent thicker CsPbBr$_{3}$ crystals. (c) Oscillation amplitude vs. frequency curve of the cantilever.  (d) Atomic force microscopy (AFM) topography of the sheet is shown, with the inset providing the corresponding height profile along the indicated line.(e and f) Morphology of the sheet in the area of 7$\mu$m $\times$ 7$\mu$m and corresponding PFM phase image respectively.}
 \end {figure}

 The as-synthesized samples yielded thin nanoplatelets of CsPbBr$_{3}$ with different lateral dimensions of the order of $\sim$ 10 $\mu$m to 50 $\mu$m as revealed by the Scanning Electron Microscopy (SEM) image shown in  Figure 1(b). For the PFM experiments, a conducting Pt/Ir coated silicon tip was brought in contact with the sample surface. The noncontact resonance frequency of the tip was 75 kHz. The contact resonance frequency of the tip was found to be 325 kHz as shown in Figure 1(c). 
\begin {figure}
\includegraphics[height=6in, width=6.5in]{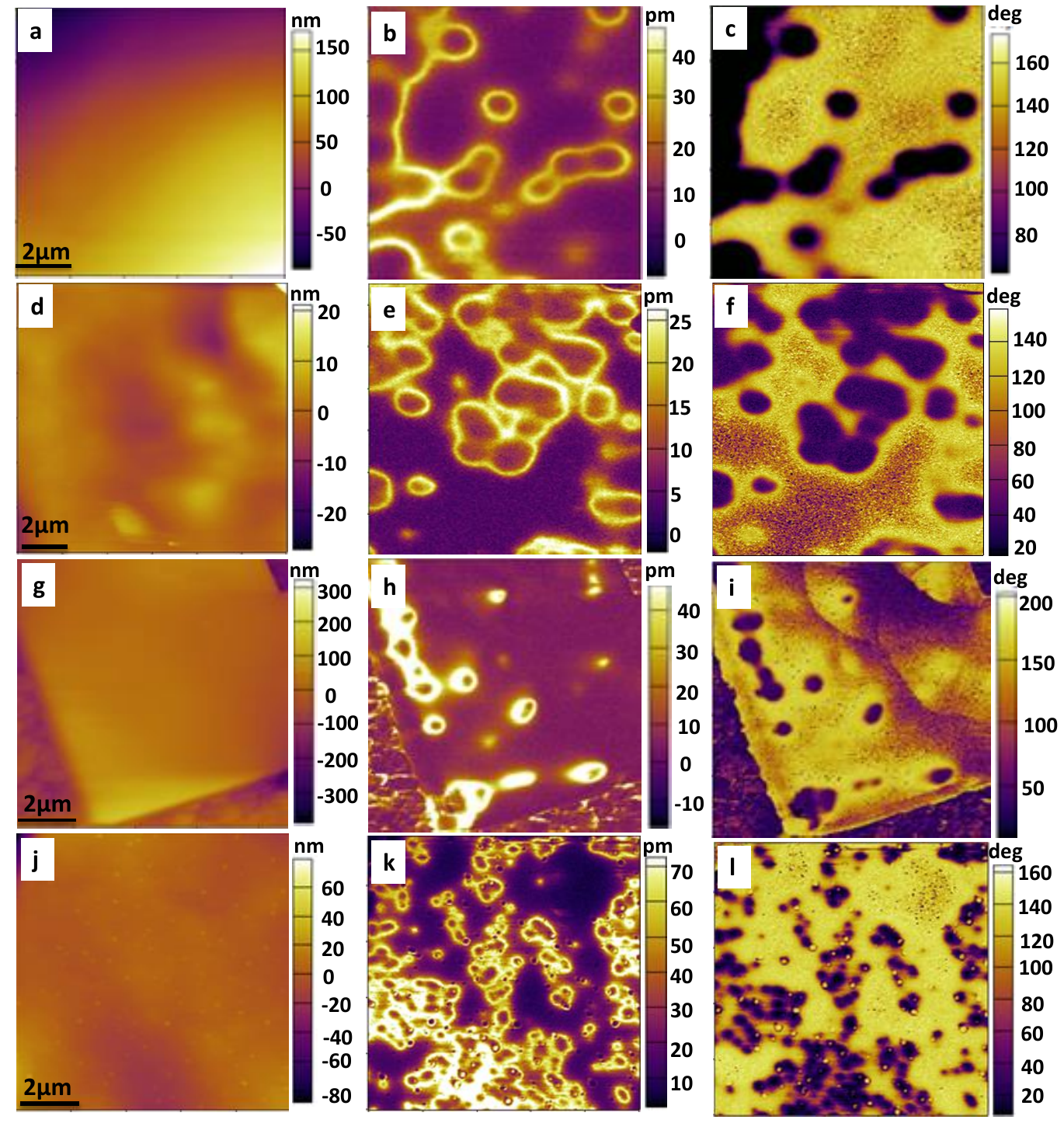}
\caption{PFM measurements on CsPbBr$_{3}$ sheets with varying thicknesses illustrate the topography, PFM amplitude, and phase images of the sheets with thickness : (a-c) 350 nm. (d-f) 300 nm (g-i) 250 nm (j-l) 125 nm}
\end {figure}
Figure 1(d) shows the surface morphology of one of the nanoplatelets and its corresponding height profile along a line (inset). It can be seen from the height profile that the thickness of the nanoplatelet is $\sim$  250nm. Figure 1(e,f) show the topography of the nanoplatelet in a smaller area of 7$\mu$m $\times$ 7$\mu$m and its corresponding PFM phase image respectively. The dark and bright contrast in the phase image, which is different from the topographic features, corresponds to oppositely polarized ferroelectric domains within the sample.
\begin {figure}
\includegraphics[height=6.8 in, width=4in]{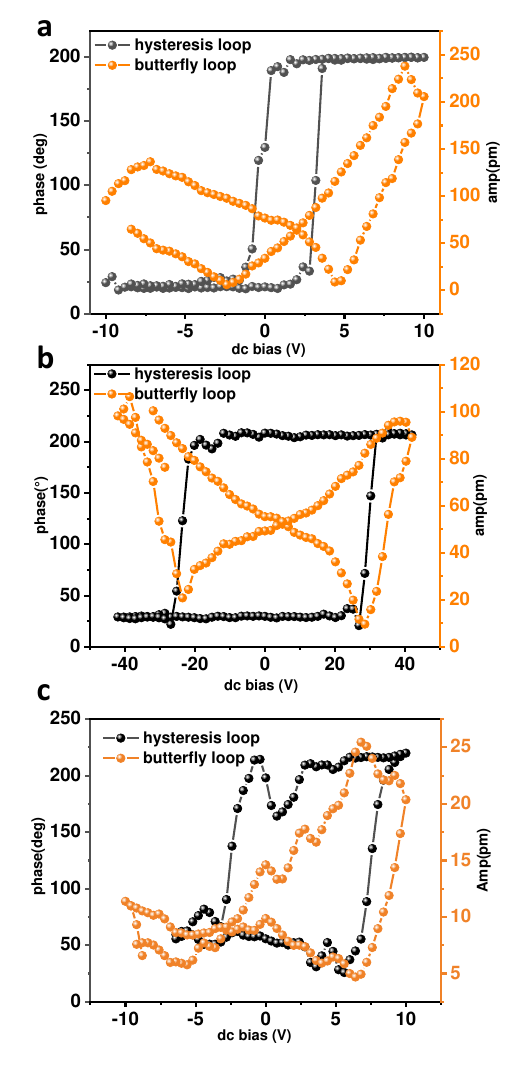}
\caption{The measured piezoelectric amplitude (amp) and phase response to applied DC bias in CsPbBr$_3$ sheets. (a) for a thickness of 150 nm, (b) 250 nm, and (c) 500 nm.}
\end {figure}
To some extent, these domains resemble magnetic bubble domains with a size on the order of microns. A characteristic contrast associated with the domain structures is typically observed in bulk and 2D ferroelectric perovskites, where thin, stripe-like, or irregularly shaped domains are usually formed with an overall fractal geometry \cite{G1., F., Jong Yeog}. In thin films and 2D materials, surface and interface effects become increasingly important, often resulting in a reduction in domain size and even the suppression of ferroelectricity below a critical thickness \cite{Weng, Fei}. 
In order to explore the effect of sheet thickness on the domain structure of quasi-2D CsPbBr$_3$, thickness dependent PFM experiments were performed. Figure 2 shows the morphology along with the corresponding PFM amplitude as well as phase images of nanoplatelets with thicknesses ranging from 350 nm to 125 nm. The morphology and the height profiles for the corresponding nanoplatelets are shown in Supplementary material Figure S1.  It is seen that the domain size decreases with the decrease in the thickness of the nanoplatelets. As discussed before, in ultrathin ferroelectrics, the depolarization field $E_d$ penalizes uniform out-of-plane polarization and favors closure-like textures whose characteristic lateral scale $w$ increases as $E_d$ decreases with thickness $t$. Therefore, these results are consistent with the previous reports showing a reduction in domain size in 2D sheets as thickness decreases \cite{Weng} and have not observed before in CsPbBr$_3$ thin nanoplatelets. The domain structures of the nanoplatelets with thicknesses of 450 nm and 175 nm are illustrated in the supplementary material (Figure S2), and the morphology of the corresponding nanoplatelets with height profile is shown in the supplementary material (Figure S3). Also, it is seen that as the thickness decreases, smaller bubble domain structures with less distinct boundaries become more prominent. This suggests that the depolarization field is significantly enhanced at reduced thicknesses. From the phase images, it is evident that the larger domains exhibit stronger contrast  
\begin {figure}
\includegraphics[height=3.5in, width=6.5in]{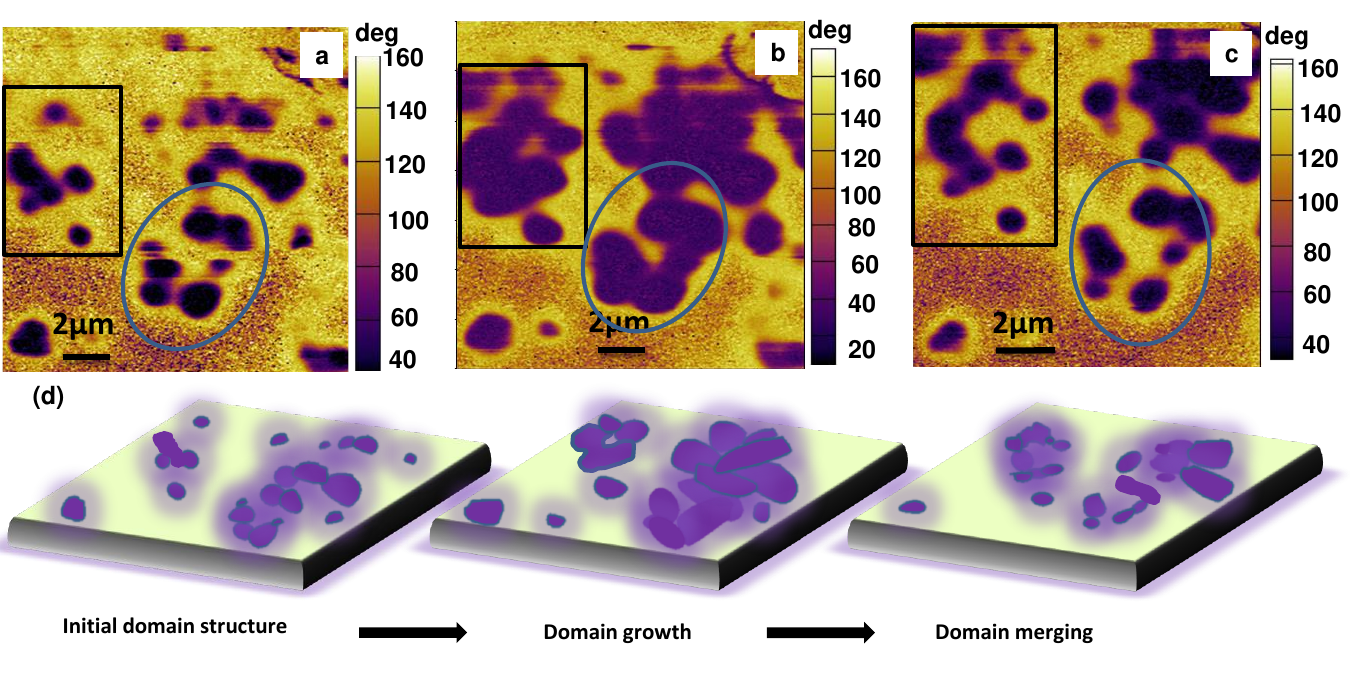}
\caption{Figure S1: PFM phase image of CsPbBr$_3$ thin sheet showing its domain evolution after successive three scans. (a) The PFM phase image of the initial scan in the area of 14$\mu$m$\times$14$\mu$m shows  the bubbles of size $\sim$ 800nm. (b) Second scan in the same area after 42 minutes. (c) shows that after two successive scans, bubble domains evolved and then transformed into a chain of bubbles that looks similar to the labyrinth pattern. (d) Schematic of the domain evolution process with successive scanning.}
\end {figure}
compared to the smaller ones, indicating that the larger domains possess higher polarization. Additionally, at lower thicknesses, two distinct domain structures are observed: conventional cylindrical domains and smaller, diffused bubbles. In Figures 2(k) and 2(l), the cylindrical domains display a clear 180$^{\circ}$ phase difference at their domain wall boundaries, characteristic of distinct up and down polarization states. In contrast, the smaller bubble-like structures exhibit less sharp boundaries, and their phase difference is also less than 180$^{\circ}$, which is a defining feature of ferroelectric bubbles.
To further confirm the ferroelectric nature of the ultrathin sheets of different thicknesses, SS-PFM experiments were performed on samples of varying thickness, as shown in Figure 3. During PFM spectroscopy, an AC-modulated DC bias was applied to the cantilever, and the local piezoelectric response was detected as the first harmonic component of the tip deflection.
The phase ($\phi$) of the electromechanical response provides information of the local polarization direction. Figure 3(a) shows that for the sheet exhibiting fuzzy bubble domains, the PFM amplitude experiences a sharp increase until reaching a maximum voltage. This rise in amplitude is attributed to an enhancement in the degree of polarization rotation. Also, it is found that the PFM amplitude with decreasing bias is at a slope markedly different from the initial rise as marked in the figure. This strongly suggests that the nanoplatelets having a small thickness and showing bubble domains of size $\sim$ 200nm also show a butterfly loop as it was seen in the bubble domains elsewhere \cite{Zhang1}. However, no such high contrast in amplitude is found in the butterfly loop (shown in Figure 3(b)) of the sheet having a domain structure different from the bubble domains. As depicted in Figures 3(a,b), the samples with thicknesses of 120 nm and 250 nm exhibit clear 180° phase switching. The orange curves in these figures correspond to the amplitude response with the applied DC bias, showing a characteristic butterfly loop, indicative of piezoelectric behavior. Additionally, we noted that the coercive voltage for the nanoplatelets having bubble-like domains is smaller than the nanoplatelets with large thicknesses. 
The sample with a thickness of approximately 500 nm (Figure 3(c)) also displayed hysteresis and a butterfly loop, similar to the sheet with a thickness of 250 nm. However, the phase and amplitude switching were significantly reduced in the thicker sheet (~500 nm).
Additionally, the absence of the ferroelectric domains in the PFM phase image shown in Supplementary Information (Figure S4 ) indicates that the sample is nonferroelectric at this thickness. 
From the SS-PFM data, the coercive field (E$_{c}$) was determined to be ~190 kV/cm and 10$^{3}$ kV/cm for the 120 nm and 250 nm thick samples, respectively. Therefore, from all our thickness-dependent scanning probe imaging and spectroscopic measurements, it is concluded that the behavior of the domain structure of thin sheets of CsPbBr$_3$ is closely similar to electric bubble domains observed in perovskite superlattices or thin films, which was not previously observed in quasi 2D  perovskite system. To further investigate the critical thickness required for the system to exhibit ferroelectric behavior, Piezoresponse Force Microscopy (PFM) measurements were conducted on a broader set of samples with varying thicknesses, as presented in the Supplementary Materials (Figures S4–S8). The PFM data (Figure S2(d)) reveal the presence of ferroelectric domains at a thickness of approximately 450 nm. However, beyond this (as indicated in Figures S4–S8), the material no longer exhibits ferroelectric behavior. Therefore, the critical thickness for the emergence of ferroelectric domain structures in this system is determined to be around 450 nm.
Since the special ferroelectric domain structures are stable only within a limited range of boundary conditions \cite{Kornev}, it is imperative to investigate the stability of the structures. To address this, we conducted a series of continuous time-dependent PFM measurements in the same area, as shown in Figure 4. The initial scan, Figure 4(a), reveals the presence of the bubble domains with indistinct boundaries. Then, repeated scanning led to the growth and merging of the bubble domains as shown in Figure 4(b). Eventually, the bubble domains in the marked (black and blue) regions transformed into a labyrinth structure as shown in Figure 4(c). This transition suggests that the stability of bubble domains is limited and the pressure imparted by the cantilever seems to be a sufficiently strong perturbation to alter them.
\begin {figure}
\includegraphics[height=3.0in, width=6.5in]{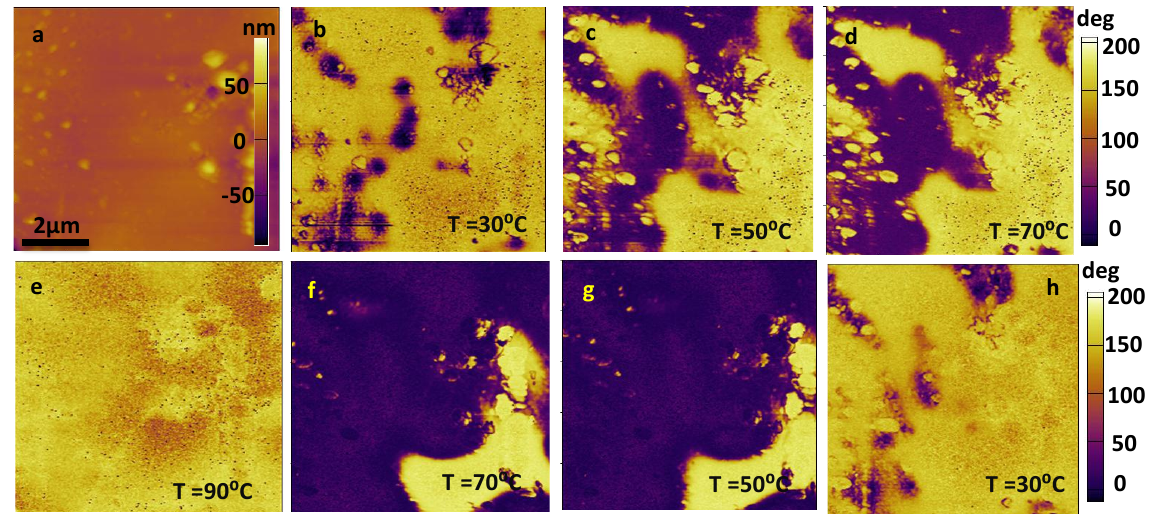}
\caption{(a) Atomic Force Microscope (AFM) image. (b)-(e) shows ferroelectric domains imaged at progressively higher temperatures, demonstrating the evolution of domain structure and indicating that the ferroelectric transition occurs below 90°C. (f)-(h) Illustrate the nucleation of domains at progressively lower temperatures.}
\end {figure}
 In ferroelectric thin films, the simplest Kittel-type scaling law predicts that the stripe or labyrinth domain period scales as $w\sim\sqrt{t}$, with additional geometric and dielectric-contrast prefactors set by the electrode/environment \cite{Igor}. As discussed before, variations in electrical boundary conditions can stabilize bubble domains, whose existence is highly sensitive to depolarization fields and screening \cite{Lai, Kornev}. Therefore, it is essential to investigate how these bubble textures evolve with temperature, which modifies both the screening efficiency and the balance between wall energy and depolarizing cost.
 We carried out temperature dependent PFM measurements on a nanoplatelet of thickness $\sim$ 250 nm, and the morphology of the sheet along with its height profile is shown in the supplementary material (Figure S5). These experiments were performed by placing the sample on a Peltier heating stage, capable of controlling the temperature from 20$^{\circ}$C to 300$^{\circ}$C. Figure 5(a, b) show the topography of the sheet and the corresponding PFM phase image at room temperature ($\sim$30$^{\circ}$C ). It is seen that a slight increase in temperature in steps of 20$^{\circ}$C led to the transformation of the bubble domains into labyrinthine structures as shown in Figure 5(b-e). At the Curie temperature ($T_c$) of $\sim$ 90°C, the labyrinth domain structures completely disappeared, indicating a phase transition from the ferroelectric phase to a paraelectric phase. The nucleation of the domain structure takes place when the temperature was decreased to room temperature ($\sim$ 30$^{\circ}$C) as shown in Figure 5 (e-h).
Bubble textures appear within an intermediate $E_d$ window where circular domns balance wall energy and depolarizing cost; they convert to labyrinths when $E_d$ is reduced (larger $t$) or under perturbations that lower the wall-curvature penalty \cite{Vivasha, Anna}. As heating reduces spontaneous polarization and wall energy, shifting this balance so that bubbles expand and merge into labyrinths before the order vanishes at $T_c$.Therefore, the bubbles were observed to evolve and merge into labyrinths before the Curie temperature of the sample.
\section*{Conclusion}
In summary, we have investigated the domain structures of CsPbBr$_{3}$ thin sheets with varying thicknesses and temperature by piezo response force microscopy. Our measurements revealed the emergence of ferroelectric domains with diverse topological textures, including bubble domains and labyrinthine domains. A significant reduction in domain size was seen as the sample thickness was decreased. At a critical thickness of $\sim$ 175 nm, an increased fraction of diffused bubble domains with indistinct boundaries was observed. Thickness-dependent PFM and SS-PFM experiments demonstrated that the topological textures vary with the nanoplatelet's thickness. Temperature-dependent PFM measurements showed that the bubble domains gradually transition into labyrinthine patterns as the temperature increases, ultimately disappearing at the Curie temperature of $\sim$ 90$^o$C, marking the phase transition from ferroelectric to paraelectric and nucleates when temperature decreased to room temperature. The observed bubble-to-labyrinth transformations and critical thickness control highlight the potential for integrating ferroelectric domain topology into halide perovskite optoelectronic devices, where domain walls could mediate enhanced photovoltaic or memory functionalities.  

\section*{Acknowledgements}
 MB acknowledges the financial support from  INSPIRE fellowship awarded by the Department of Science and Technology (DST), Government of India.  GS acknowledges financial assistance from Science and Engineering Research Board (SERB), Govt. of India (grant number: \textbf{CRG/2021/006395}). AR acknowledges funding support from DST SERB Grant No. CRG/2021/005659.
\section*{Data availability}
The data that support the findings of this study are available within the article and the supplementary file.
\section*{Author Declarations } 
\subsection*{Competing interests}
The authors declare no competing financial interests.

\section*{Supplementary Materials} Thickness Analysis using AFM. Additional PFM measurements for critical thickness analysis. Morphology of the sample along with the height profile of the nanoplatelet on which thickness and temperature-dependent PFM experiments were performed.
\subsection*{Nanoplatelet Thickness Analysis Using AFM}

\renewcommand{\thefigure}{S1} 
\begin {figure}[ht!]
\begin{center}
\includegraphics[height=6in, width=6in]{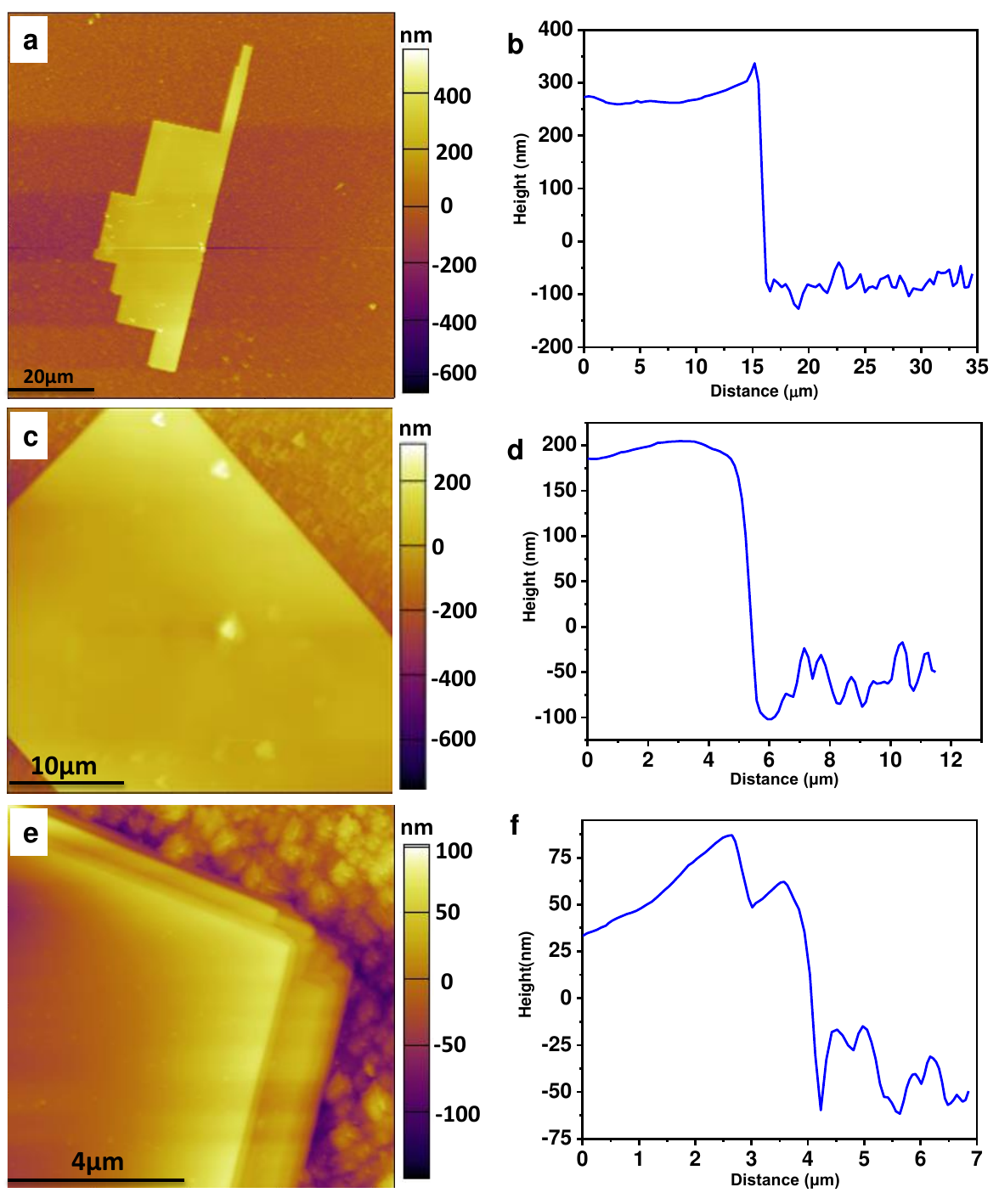}
\end{center}
\caption{ Morphology (Left )and corresponding height profile of CsPbBr$_3$ nanoplatelets with thicknesses (a,b) 350 nm, (c,d) 300 nm, and (e,f) 125 nm.}
\end {figure}
\renewcommand{\thefigure}{\arabic{figure}}
\subsection*{Additional PFM measurements for critical thickness analysis}
In Figure S2(a), we show the topography of the nanoplatelet with a scan size of 10$\mu$m $\times$10$\mu$m for a 175 nm thick sample. PFM phase image shows the ferroelectric bubble domains of size ~ 500 nm (Figure S2(b)). Figure S1(c) shows the topography of different nanoplatelet with a thickness of 450 nm. The large domain size in the phase image (Figure S2(d)) indicates that the domain size increases with the thickness of the sample. The AFM images, along with the height profiles, are shown in Figure S3.
\renewcommand{\thefigure}{S2}
\begin{figure}[ht!]
\begin{center}    
\includegraphics[height=5in, width=6in]{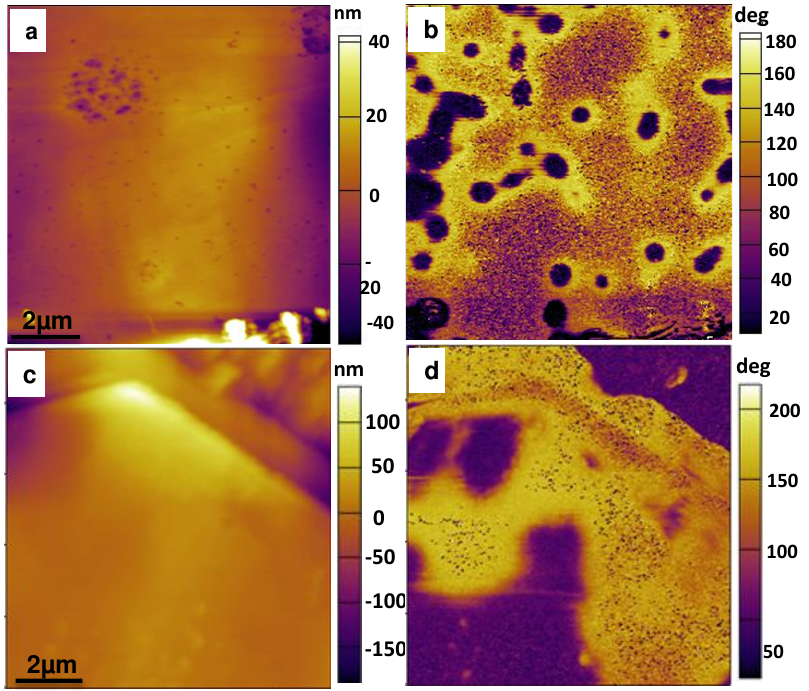}
\end{center}
\caption{(a,c) Topography of the nanoplatelets for the thickness of 175 nm and 450 nm, respectively, and corresponding PFM phase images (b, d) showing the ferroelectric bubble domains.}
\end {figure}
\renewcommand{\thefigure}{\arabic{figure}}

\renewcommand{\thefigure}{S3}
\begin {figure}[ht!]
\begin{center} 
\includegraphics[height=5in, width=6in]{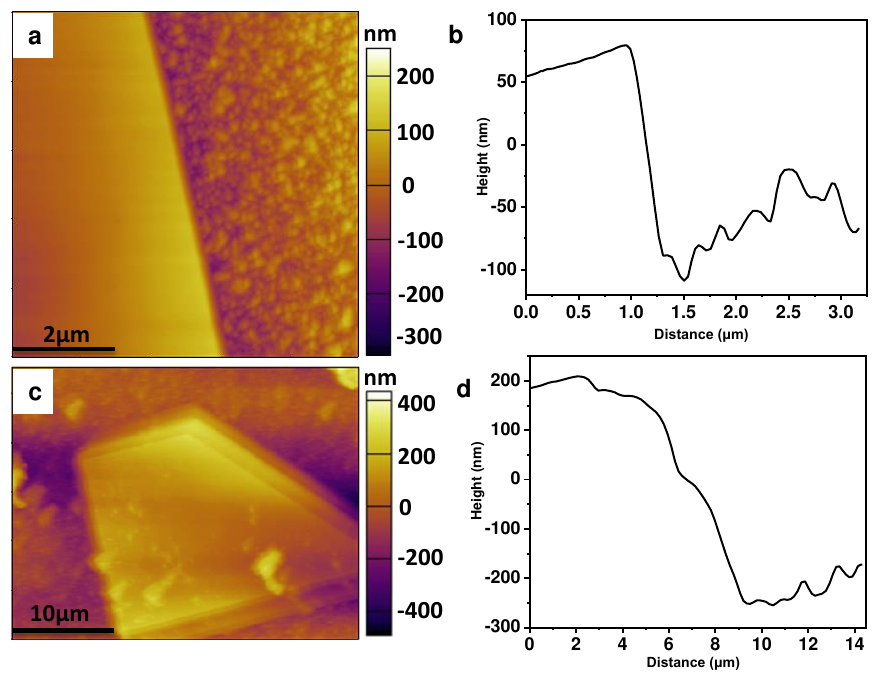}
\end{center} 
\caption{(a,c) Topography and (b, d) corresponding height profile of the nanoplatelets showing the thickness of 175 nm and 450 nm, respectively.}
\end {figure}
\renewcommand{\thefigure}{\arabic{figure}}
\renewcommand{\thefigure}{S4}
\begin {figure}[ht!]
\begin{center}   
\includegraphics[height=5in, width=6in]{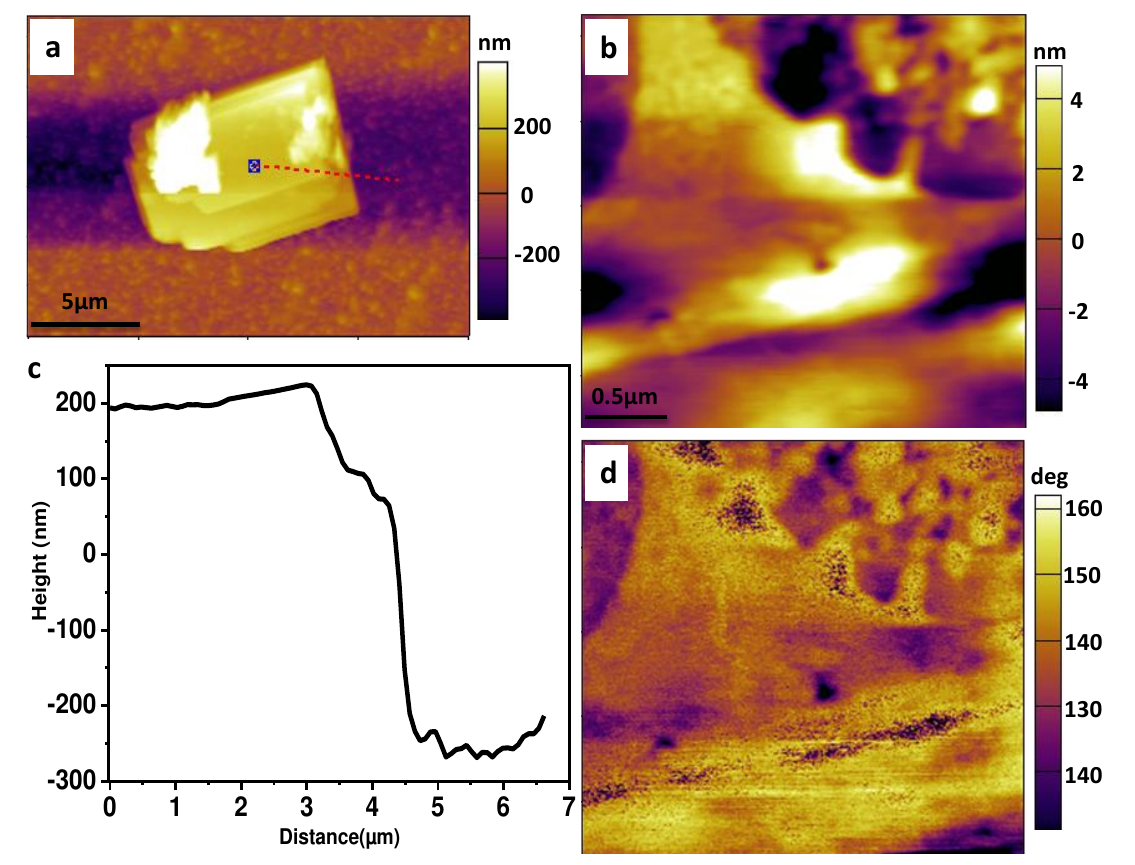 }
\end{center} 
\caption{(a) AFM image of a nanoplatelet. (b)  The corresponding height profile of the nanoplatelet along the red line shows that the thickness of the nanoplatelet is $\sim$ 465 nm. (c) Topographic image of the sample in the small area. (d) The corresponding PFM phase shows no ferroelectric domains.}
\end {figure}
\renewcommand{\thefigure}{\arabic{figure}}

\renewcommand{\thefigure}{S5}
\begin {figure}[ht!]]
\begin{center} 
\includegraphics[height=5in, width=6in]{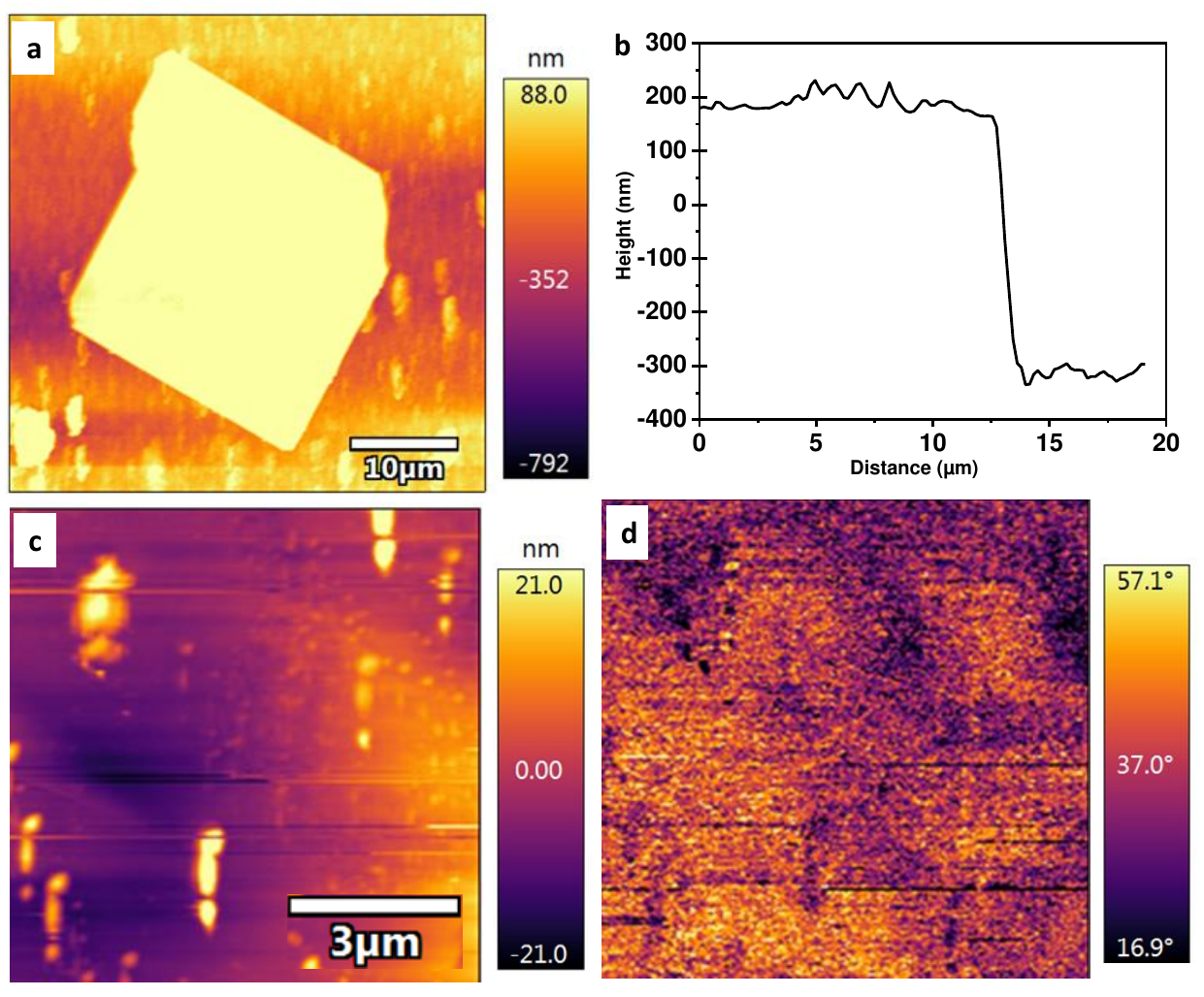 }
\end{center} 
\caption{(a,b) AFM image of nanoplatelet and the corresponding height profile of the sheet along the red line shows that the thickness of the nanoplatelet is $\sim$500  nm. (c,d) The topographic image of the sample in the small area and the corresponding PFM phase shows no ferroelectric domains.}
\end {figure}
\renewcommand{\thefigure}{\arabic{figure}}
\renewcommand{\thefigure}{S6}
\begin {figure}[ht!]
\begin{center} 
\includegraphics[height=4in, width=5.8in]{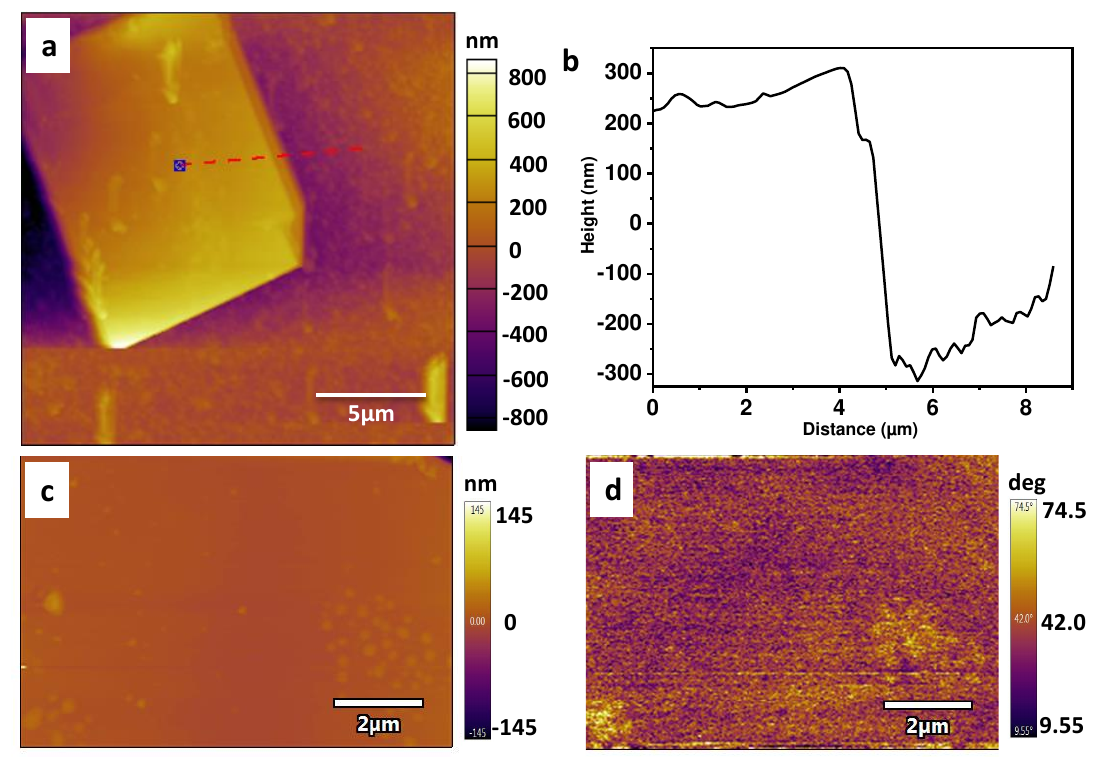 }
\end{center} 
\caption{(a) AFM image of a nanoplatelet in an area of 18$\mu$m $\times$ 18$\mu$m. (b)  The height profile of the sheet along the red line shows that the thickness of the nanoplatelet is $\sim$ 550 nm. (c) AFM image of the sample in a small area of 10$\mu$m $\times$10$\mu$m. (d) The corresponding PFM phase shows no ferroelectric domains.}
\end {figure}
\renewcommand{\thefigure}{\arabic{figure}}

\renewcommand{\thefigure}{S7}
\begin {figure}[ht!]
\begin{center} 
\includegraphics[height=4.5in, width=5.5in]{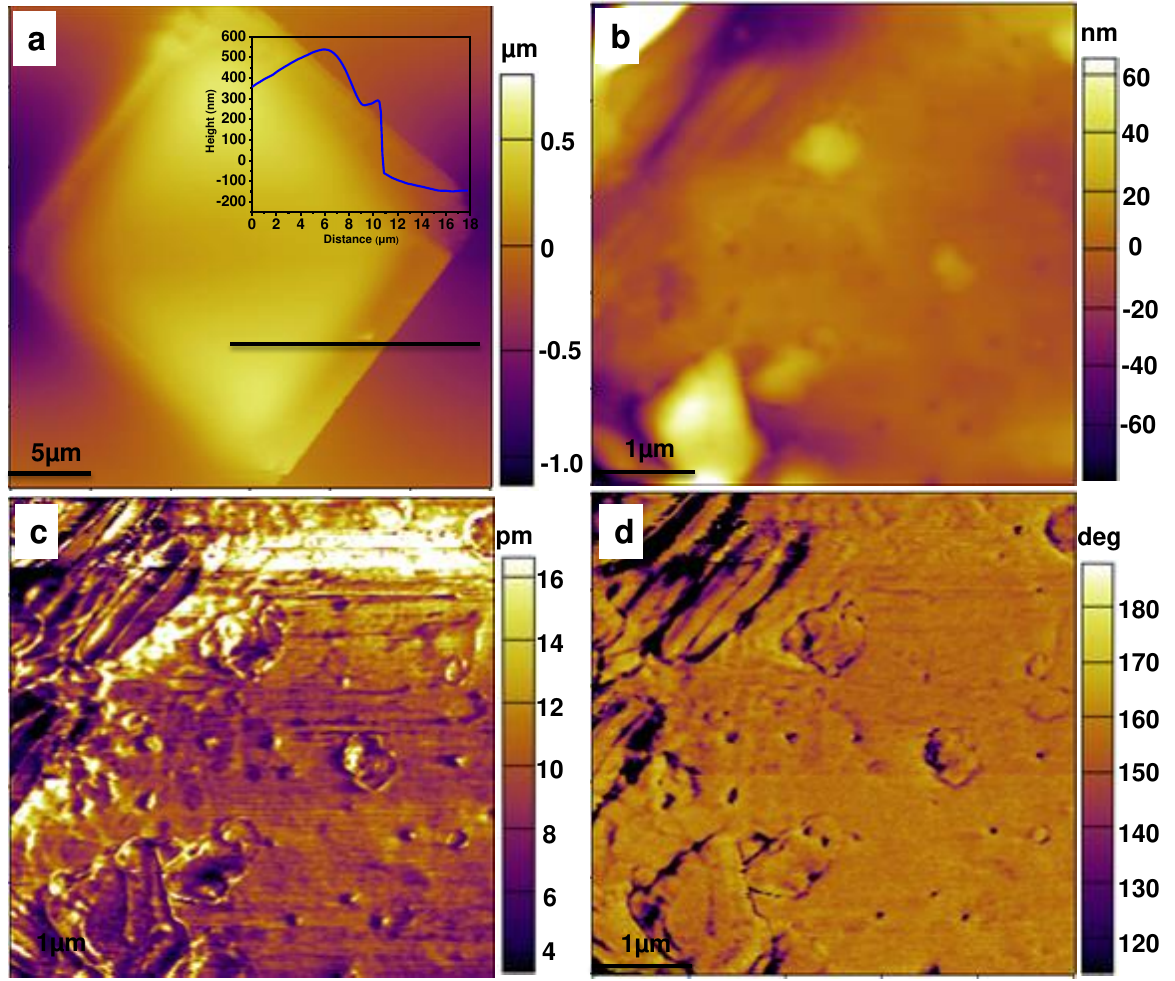}
\end{center} 
\caption{(a)AFM topography of the nanoplatelet. The inset shows the corresponding height profile along the black line. (b) Morphology of the nanoplatelet in the area of 5$\mu$m $\times$5$\mu$m and (c$\&$d) corresponding PFM phase image show that no ferroelectric domains are observed,  on the nanoplatelet of thickness $\sim$ 600 nm as shown in (a).}
\end {figure}
\renewcommand{\thefigure}{\arabic{figure}}
\renewcommand{\thefigure}{S8}
\begin {figure}[ht!]
\begin{center} 
\includegraphics[height=4.3in, width=5.8in]{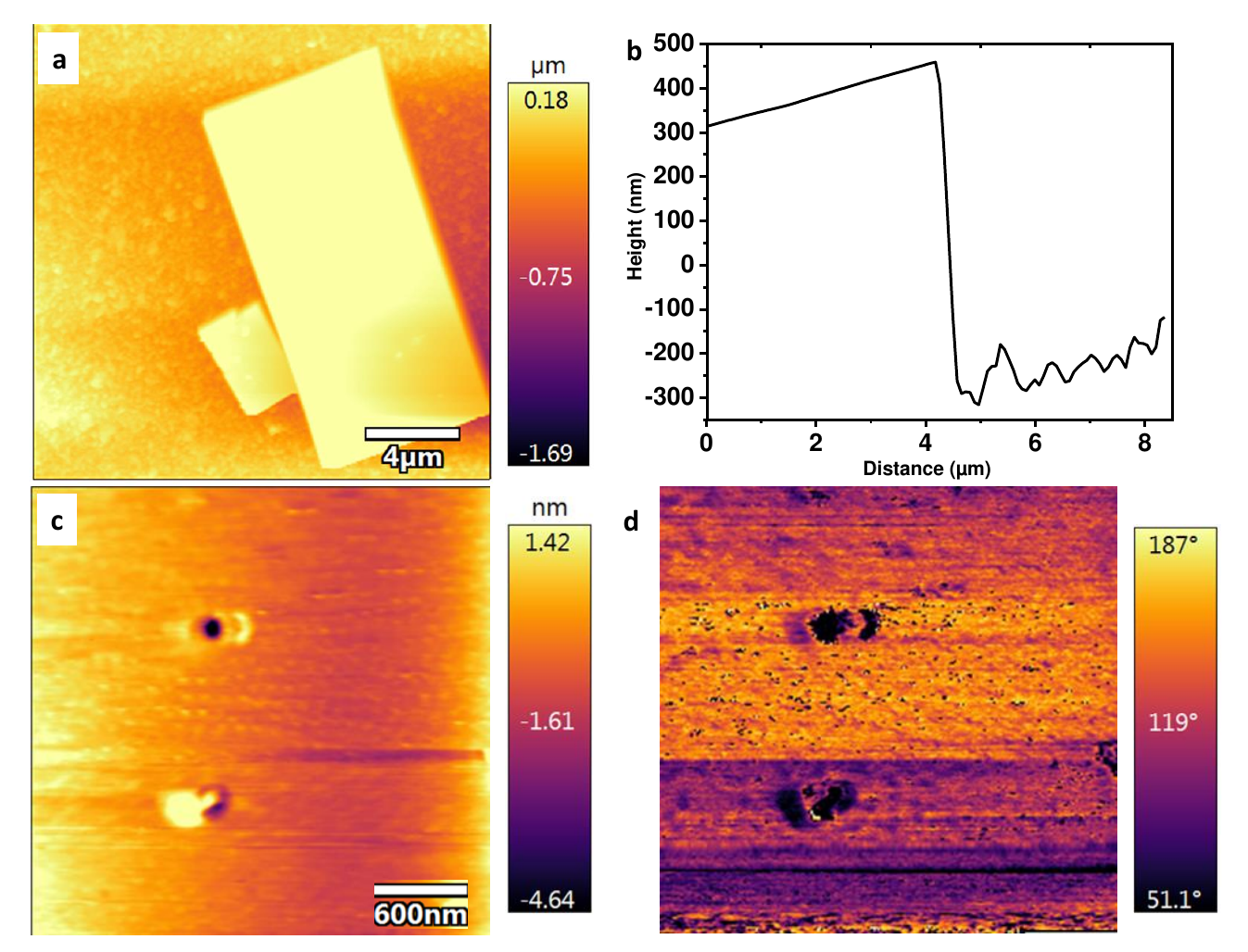}
\end{center} 
\caption{(a, b) AFM image of the nanoplatelet in a large area of 30 µm × 30 µm and corresponding height profile shows the thickness of the sample $\sim$ 650 nm (c, d ) Topography of the nanoplatelet in the area of 2.5 µm × 2.5 µm. (d) The corresponding PFM phase image shows the absence of ferroelectric domains.}
\end {figure}
\renewcommand{\thefigure}{\arabic{figure}}



\renewcommand{\thefigure}{S9}
\begin {figure}[ht!]
\begin{center} 
\includegraphics[height=4.5in, width=5.5in]{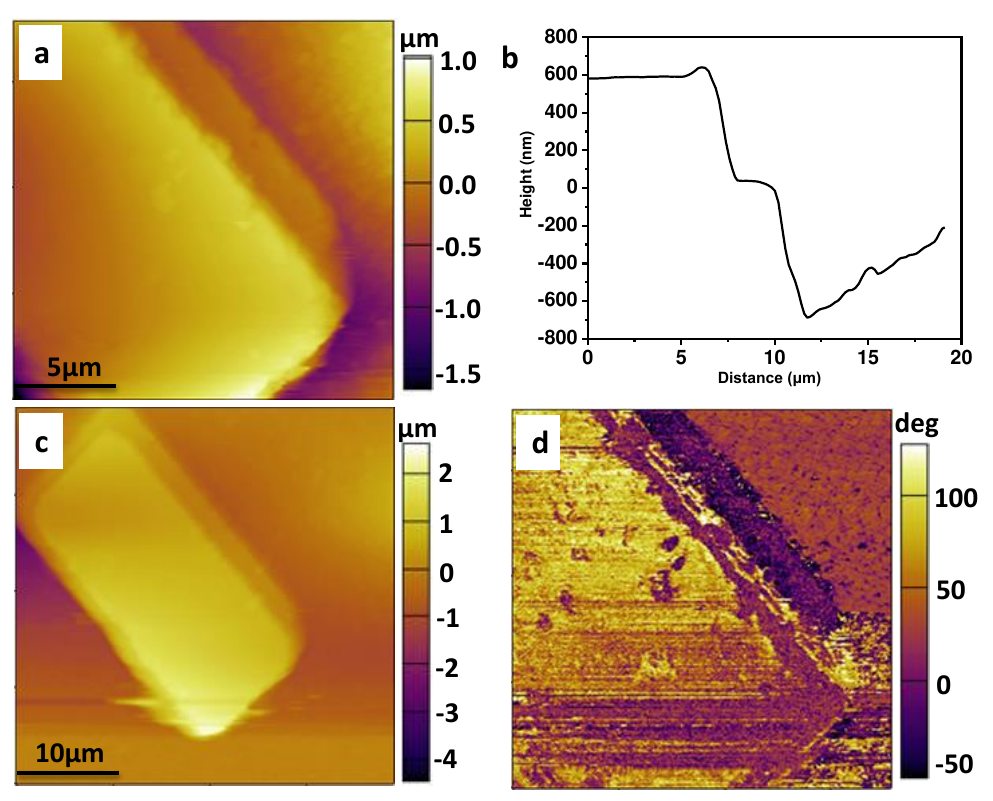}
\end{center} 
\caption{(a, b) AFM image of the nanoplatelate in the large area of 30 µm × 30 µm and the corresponding height profile shows the thickness of the sample $\sim$ 1.3 µm. (c, d ) The magnified AFM image shows the topography of the nanoplatelet. (d) The corresponding PFM phase image shows the absence of ferroelectric domains.}
\end {figure}
\renewcommand{\thefigure}{\arabic{figure}}
\subsection*{Surface Potential Mapping via KPFM}
\renewcommand{\thefigure}{S10}
\begin{figure}[H]
\centering  
\includegraphics[height=4.56in, width=6in]{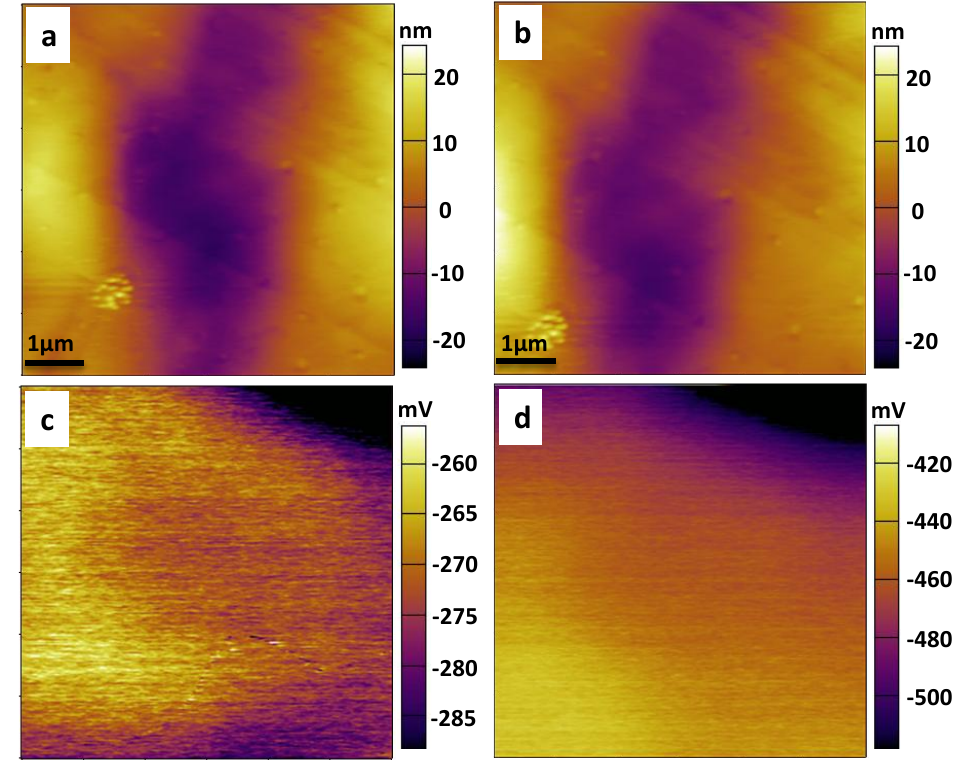 }

\caption{(a) Topographic image in an area of 6 µm × 6 µm. (c) The corresponding surface potential map in the absence of light shows the average contact surface potential difference of approximately 270 mV. (b) Topographic image of the nanoplatelet with illumination of light. (d) Corresponding surface potential map showing the modulation of the surface potential with the illumination of light.}
\end {figure}
\renewcommand{\thefigure}{\arabic{figure}}
 Kelvin Probe Force Microscopy (KPFM) \cite{Lee} is a powerful and widely used characterization technique for investigating electronic properties of organic, inorganic, and hybrid photovoltaic materials \cite{ Fuchs, Xiao}. It provides spatially resolved information on key parameters such as surface potential, local surface photovoltage, and work function in photoactive thin films and device architectures.  In this study,to understand the enhanced photovoltaic behavior of the sample, KPFM measurements were carried out. These measurements were performed in non-contact mode, with the sample surface electrically grounded and an AC-modulated DC bias applied to the conductive AFM tip. Figure S10 presents both the topography and the corresponding surface potential maps of the nanoplatelet under dark conditions and upon illumination.\\
 In Figure 10(c), dark and bright contrast shows the contact potential difference (CPD) corresponding to the up and down polarised domains respectively. Upon illumination, a notable increase in CPD (shown in Figure 10(d)) indicates the interaction between photogenerated charge carriers and the internal polarization fields within the material. The intrinsic electric field associated with each domain type—up-polarized (polarization vector pointing toward the surface) and down-polarized (polarization pointing into the bulk)—drives the separation of electron–hole pairs differently. In up-polarized domains, electrons tend to accumulate at the surface, lowering the local work function and making the surface potential more negative. In contrast, down-polarized domains tend to accumulate holes at the surface, which raises the local work function, though in the current observation, the net effect still results in an overall negative shift in surface potential across both domains. The negative values in the potential map's scale bar indicate that the sample possesses a higher work function compared to the AFM tip.
 Highly Oriented Pyrolytic Graphite (HOPG) was employed as a reference material \cite{Banoo} to calibrate the AFM tip for work function determination, as shown in Figure S11.
 The work function of the sample was calculated using the following relationship:
$\phi_{tip}$- $\phi_{sample}$ = V$_{CPD}$
where $\phi_{sample}$, $\phi_{tip}$ is the work function of the sample and the tip, respectively, and V$_{CPD}$ is the difference in the contact potential between them..

 \renewcommand{\thefigure}{S11}
\begin {figure}[ht!]
\begin{center} 
\includegraphics[height=2.7in, width=6in]{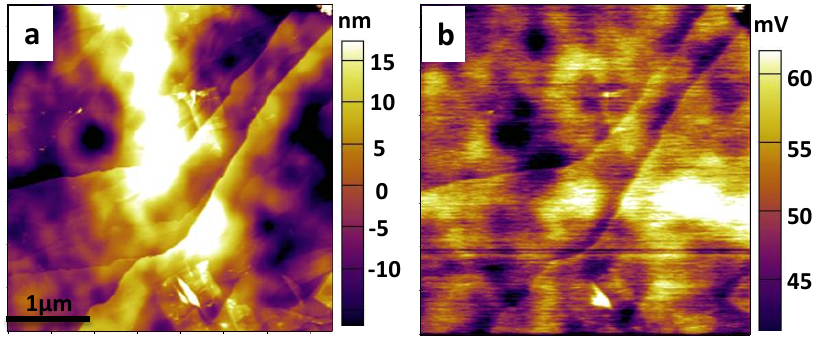}
\end{center} 
\caption{(a) Topography and (b) surface potential map of Highly Ordered Pyrolytic Graphite (HOPG), used for tip calibration. HOPG has a work function $\Phi$ of 4.66 eV and a contact potential difference V$_{CPD}$ of 52 mV. }
\end {figure}
\renewcommand{\thefigure}{\arabic{figure}}
 \renewcommand{\thefigure}{S12}
\begin {figure}[ht!]
\begin{center} 
\includegraphics[height=2.7in, width=6in]{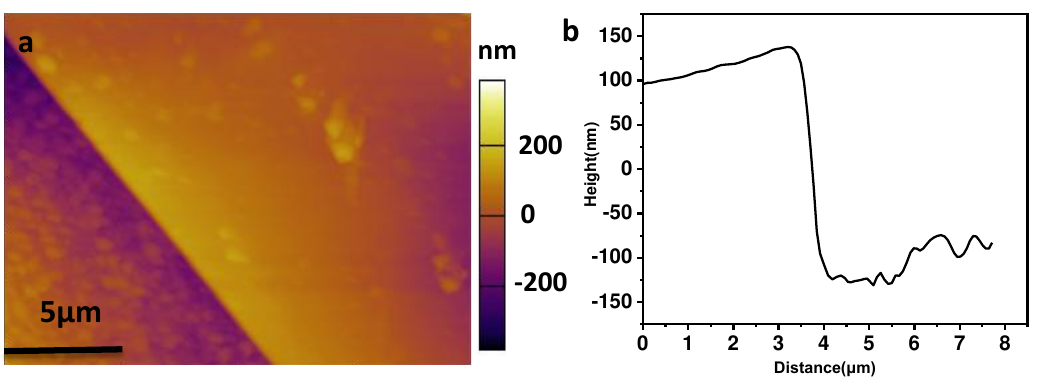} 
\end{center} 
\caption{(a) Topography of nanoplatelet in the area of 20$\mu$m$\times$20$\mu$m, the violet color shows the ITO-coated glass surface and the slanted yellow is the sample surface. (b) The corresponding height profile of the nanoplatelet shows that the thickness of the sheet is 250nm.}
\end {figure}
\renewcommand{\thefigure}{\arabic{figure}}
\pagebreak

\end{document}